\documentclass[10pt,twocolumn,showpacs,aps,prd,preprintnumbers,amsmath,amssymb,superscriptaddress,nofootinbib,postscript,showkeys]{revtex4-1}
\usepackage{enumerate}
\usepackage{epsfig} 
\usepackage{float} 
\usepackage[caption = false]{subfig}
\usepackage{wasysym} 
\usepackage{mathrsfs} 
\usepackage{amsfonts} 
\usepackage{amsbsy} 
\usepackage{amscd} 
\usepackage{pstricks} 
\usepackage{multirow} 
\usepackage{tikz}
\usepackage{color}
\usepackage{slashed}
\usepackage{array}
\usepackage{tablefootnote} 
\usepackage{enumitem}
\usepackage[utf8]{inputenc}
\usepackage{hyperref}
\usepackage{color}
\definecolor{myred}{rgb}{0.6,0,0} %usage:  {\textcolor{myred}{Hello World}}
\definecolor{myblue}{rgb}{0,0.2,0.4}
\definecolor{mygreen}{rgb}{0,0.9,0.1}
\definecolor{hc}{rgb}{.9,0.1,0.7}
\definecolor{hcout}{rgb}{.9,0.7,0.9}
\definecolor{Orange}{rgb}{1.,0.65,0.}
\makeatletter

\newcommand{\fmslash}[2][0mu]{%
  \mathchoice
    {\fmsl@sh\displaystyle{#1}{#2}}%
    {\fmsl@sh\textstyle{#1}{#2}}%
    {\fmsl@sh\scriptstyle{#1}{#2}}%
    {\fmsl@sh\scriptscriptstyle{#1}{#2}}}
\newcommand{\fmsl@sh}[3]{%
  \m@th\ooalign{$\hfil#1\mkern#2/\hfil$\crcr$#1#3$}}
\makeatother

\newcommand{\lsim}{{\;\raise0.3ex\hbox{$<$\kern-0.75em\raise-1.1ex\hbox{$\sim$}}\;}}
\newcommand{\gsim}{{\;\raise0.3ex\hbox{$>$\kern-0.75em\raise-1.1ex\hbox{$\sim$}}\;}}

\usepackage{array}
\newcolumntype{C}[1]{>{\centering\arraybackslash$}p{#1}<{$}}

\usepackage{bm} % bold in math mode
\newcommand{\be}{\begin{equation}}
\newcommand{\ee}{\end{equation}}
\newcommand{\bes}{\begin{equation*}}
\newcommand{\ees}{\end{equation*}}
\newcommand{\bea}{\begin{eqnarray}}
\newcommand{\eea}{\end{eqnarray}}
\newcommand{\beas}{\begin{eqnarray*}}
\newcommand{\eeas}{\end{eqnarray*}}

\renewcommand{\labelitemi}{}
\begin{document}
\title{Decaying fermionic warm dark matter and XENON1T electronic recoil excess} 

\author{Koushik Dutta}
\email{koushik@iiserkol.ac.in } 
\affiliation{Department of Physical Sciences, 
Indian Institute of Science Education and Research, 
Kolkata, Mohanpur - 741246, India}
\author{Avirup Ghosh}
\email{avirupghosh@hri.res.in}
\affiliation{Regional Centre for Accelerator-based Particle Physics,
Harish-Chandra Research Institute, HBNI,
Chhatnag Road, Jhunsi, Allahabad - 211 019, India} 
\author{Arpan Kar} 
\email{arpankar@hri.res.in}
\affiliation{Regional Centre for Accelerator-based Particle Physics,
Harish-Chandra Research Institute, HBNI,
Chhatnag Road, Jhunsi, Allahabad - 211 019, India} 
\author{Biswarup Mukhopadhyaya}
\email{biswarup@iiserkol.ac.in} 
\affiliation{Department of Physical Sciences, 
Indian Institute of Science Education and Research, 
Kolkata, Mohanpur - 741246, India} 

%\date{\today}
\preprint{HRI-RECAPP-2021-004}

\begin{abstract}
In the light of the recently observed XENON1T electronic recoil (ER) data, we investigate the
possibility of constraining the parameter space of a generic fermionic warm dark matter (WDM),
decaying into a standard model (SM) neutrino and a photon. The photon as a decay product,
when produced inside the XENON1T chamber, interacts with an electron of a xenon (Xe) atom,
leading to a contribution in the observed ER data. 
We add this dark matter (DM) induced signal
over the standard background ($\rm B_0$) considered by the 
XENON1T collaboration and perform a 
$\chi^2$ fit against the XENON1T data to obtain the best-fit values of the DM decay width and
the associated $95\%$ confidence level (C.L.) band for DM mass ($m_\chi$) varied in the range $2 - 60$ keV.
Additionally, we have extended our analysis by including %for 
two other background models available in the literature and in 
each case, the corresponding limits on the DM decay width
are estimated for DM mass ($m_\chi$) in the domain $2 - 18$ keV. By comparing the constraints, obtained
by fitting the XENON1T data, with the upper limits arising from various existing astrophysical and
cosmological observations, we find that, for the background model $\rm B_0$, a fair amount of the DM
parameter space is allowed at $95\%$ C.L. for DM masses outside the range $3.5\,{\rm keV} \lesssim m_\chi \lesssim 8.5\,{\rm keV}$.
However, in case of other two background models, reasonable parts of the DM parameter space are
favoured at $95\%$ C.L. by all astrophysical data for all DM masses in the range $2 - 18$ keV.
%
%However, unlike $\rm B_0$, in case of other two background models, reasonable parts of the DM parameter space are
%favoured at $95\%$ C.L. by all astrophysical data \blu{for all DM masses in the range $2 - 18$ keV.
%----------------------------------------------------------
%\blu{
%We add this dark matter (DM) induced signal over four distinct backgrounds (taking one at a time) and 
%perform a single parameter $\chi^2$ fit against the XENON1T data to obtain the best-fit values of the 
%DM decay width and the associated 95$\%$ confidence level (C.L.) bands for DM mass varied in the range $2 - 18$ keV.
%By comparing the constraints, obtained by fitting the %XENON1T data, with the upper limits, arising from various 
%existing astrophysical and cosmological observations, we find %that a fair amount of the DM parameter space is 
%allowed at $95\%$ %C.L., for each of the background models considered.} 
%----------------------------------------------------------
\end{abstract}

%==============================================================================
% {Comparing these results with the constraints obtained from astrophysical 
% and cosmological observations we find that such decaying DM scenario is 
% allowed at $95\%$ confidence level (C.L.) for DM mass in the range $2 - 18$ keV for three 
% of the considered background models, barring the 
% standard background (${\rm B_0}$) consiedred by XENON1T collaboration
% Comparing these results with the constraints obtained from 
% astrophysical and cosmological observations we found that if one considers the 95$\%$ confidence 
% level (C.L.) band containing the best-fit decay width, the DM  
% signal is allowed for the mass range $2 - 18$ keV, barring the 
% standard background ${\rm B_0}$. While for the standard background 
% (${\rm B_0}$), such signal is consistent with all observational 
% constraints for DM masses $\lesssim 3$ keV and  $\gtrsim 8$ keV.}
%==============================================================================

%\\$\textrm{}$\hfill \today}
%\preprint{\today}


%\pacs{14.80.Ly,12.60.Jv, 95.60.Cq} 
\maketitle

%\newpage

%******************************************************* 
\section{Introduction}
\label{sec:intro}

Recently, XENON1T has reported an excess in its electronic recoil (ER) events in the energy range 
$2 - 3$ keV~\cite{Aprile:2020tmw,xenon_collaboration_2020_4273099}.
Several scenarios have already been proposed in explaining this excess as potential signals of bosonic dark matter (DM) candidates
~\cite{Kannike:2020agf,Takahashi:2020bpq,Alonso-Alvarez:2020cdv,Buch:2020mrg,Nakayama:2020ikz,An:2020bxd,Bloch:2020uzh,Lindner:2020kko,Okada:2020evk}, 
Primakoff process involving solar axions~\cite{Athron:2020maw,Gao:2020wer,Dent:2020jhf,Li:2020naa,Cacciapaglia:2020kbf}, 
neutrino magnetic moment~\cite{Shakeri:2020wvk,Shoemaker:2020kji,Miranda:2020kwy,Khan:2020vaf,Babu:2020ivd} %etc.
along with many other new physics possibilities~\cite{Fornal:2020npv,Bally:2020yid,Xu:2020qsy}.
In~\cite{Choi:2020udy}, a decaying fermionic warm dark matter (WDM) particle, which produces a dark photon,
has been considered as a possible explanation. 
In most of these cases, the interactions which cause the XENON1T ER events are difficult to probe in indirect search observations. 
Some previous works~\cite{Bhattacherjee:2020qmv,Szydagis:2020isq,Robinson:2020gfu}
have also tried to fit the observed data using various possible background models 
(other than the standard background $\rm B_0$~\cite{Aprile:2020tmw,xenon_collaboration_2020_4273099}), 
leaving some small scopes for adding new signals. 
 
In this paper, we assume a fermionic DM ($\chi$) whose mass 
($m_{\chi}$) is in the keV range and which decays into a photon ($\gamma$) along with a SM neutrino ($\nu$). 
It is a scenario studied in the past in several 
contexts~\cite{Asaka:2005pn,Asaka:2006nq,Hofmann:2016urz,Dey:2011zd,Takayama:2000uz,Covi:2009pq,Kong:2014gea,Choi:2014tva,Giudice:1998bp,Kolda:2014ppa,Bomark:2014yja}. 
One theoretical basis of such a possibility is offered by R-parity 
violating supersymmetry (SUSY) with a keV-scale warm dark matter 
candidate such as the gravitino or the 
axino~\cite{Dey:2011zd,Takayama:2000uz,Covi:2009pq,Kong:2014gea,Choi:2014tva,Giudice:1998bp,Kolda:2014ppa,Bomark:2014yja}. 
The breakdown of R-parity (defined as $R = (-1)^{3B+L+2J}$) via bilinear L-breaking terms in the 
superpotential leads to photino-neutrino mixing. This in turn  can 
drive the decay of the gravitino/axino into a photon and a 
neutrino, with the photon in the right energy range answering  to 
the XENON1T observation. Keeping this in mind, we perform 
a model-independent analysis, assuming tentatively that the keV-scale DM particle saturates the observed 
DM densities in the galaxies.

%nevertheless
%One theoretical basis for such a possibility is offered by R-parity violating supersymmetry (SUSY) with a keV-scale warm dark matter 
%candidate such as the gravitino ($\tilde{G}$)~\cite{Takayama:2000uz} %%\cite{Giudice:1998bp}. 
%The breakdown of R-parity (defined as $R=(-1)^{3B+L+2J}$) via bilinear 
%L-breaking terms in the superpotential leads to photino-neutrino mixing. This in turn can drive the decay $\tilde{G}\rightarrow \nu \gamma$, 
%with the photon in the right energy range answering to the XENON1T observation. Similar expectations arise in the case of a keV-scale axino, 
%too~\cite{Covi:2009pq}. This kind of dark matter decay influencing the XENON1T data may thus be treated in a model-independent fashion, 
%assuming tentatively that the keV-range dark matter particle saturates the observed DM densities in galaxies.

%----------------------------------------------------------
%~\footnote{In principle, decays of DM outside the XENON1T chamber can also impart energy to the electrons of the Xe atoms. \blu{While the photon in such cases gets mostly shielded, we have verified that the weak interaction effect from neutrinos is down by several orders of magnitude compared to what DM decays within the detector can do with the help of photons, as discussed in our work.}}. 
%----------------------------------------------------------
In this scenario, the local population of $\chi$ \textit{decays inside} 
the XENON1T detector and the resulting photons contribute to 
the XENON1T ER events by means of photoelectric effect.
We take into account the standard background model ($\rm B_0$)
(as considered in \cite{Aprile:2020tmw,xenon_collaboration_2020_4273099}) 
and carry out a likelihood analysis to fit the DM induced signal,
originating from the DM decays (i.e., $\chi \rightarrow \gamma \nu$) within the XENON1T detector, 
against the ER data~\cite{Aprile:2020tmw,xenon_collaboration_2020_4273099}. 
The resulting best-fit values of the DM decay width 
$\Gamma$ and the corresponding $95\%$ confidence level (C.L.) 
band are obtained as functions of
the DM mass $m_{\chi}$ which varies in the range $2 - 60$ keV.
For illustration, a similar analysis is performed 
for two other background models, studied 
in~\cite{Szydagis:2020isq}, and in each case the corresponding 
constraints on the DM parameter space are derived.
%----------------------------------------------------------
%\blu{
%We take into account four different background models~\cite{Aprile:2020tmw,Szydagis:2020isq} 
%and carry out a likelihood analysis to fit the DM induced signal against the ER data.
%The resulting best-fit values of the DM decay width $\Gamma$ 
%and the corresponding $95\%$ confidence level (C.L.) bands are obtained as a function of the DM mass $m_{\chi}$ 
%which varies in the range $2 - 18$ keV. }
%----------------------------------------------------------

In addition, decays of DM particles $\chi$ 
in galactic and extra-galactic sources
produce X-ray photons which can be detected in various space-based 
X-ray telescopes such as Chandra, NuSTAR, XMM-Newton, HEAO-1 etc. 
Non-detections of any new signal in these experiments result in 
the upper limits in the $\Gamma - m_{\chi}$ 
plane~\cite{Watson:2011dw,Sicilian:2020glg,Perez:2016tcq,Ng:2019gch,Jeltema:2008ax,Boyarsky:2005us,Essig:2013goa}, 
among which, Chandra and NuSTAR observations provide the strongest constraints.
Simultaneously, the constraint coming from the CMB observation of 
Planck has also been taken into consideration~\cite{Ade:2015xua,Oldengott:2016yjc}.  
We compare our results with the aforementioned 
observational constraints and show that, 
for the standard background model $\rm B_0$, some parts of the DM parameter space in the $95\%$ C.L. region, 
obtained by fitting the XENON1T data, are consistent with all astrophysical observations.
In case of the two additional backgrounds, too,  
substantial portions of the DM parameter space 
in the $95\%$ C.L. bands are allowed by all observations.

The paper is organized as follows : In Sec.~\ref{sec:model}, we 
discuss the DM model, backgrounds and corresponding analysis methodology. 
In Sec.~\ref{sec:cons}, we list up the possible astrophysical and cosmological 
observations that can constrain the parameter space of keV DM.
In Sec.~\ref{sec:results}, we present our results and compare them with the observational constraints. 
We finally conclude in Sec.~\ref{sec:conclusion}.           

\section{Model and analysis}
\label{sec:model}

In order to see the implications of the recently observed XENON1T electronic recoil (ER) data on a new physics signal,
we consider a scenario consisting of a fermionic dark matter (DM) $\chi$, decaying into a SM neutrino ($\nu$) and a 
photon ($\gamma$). If the DM rest mass ($m_\chi$) is in the range $\mathcal{O}(\rm keV)$ (as considered in this work), the 
corresponding energy of the produced photon will be $E_\gamma \simeq m_\chi/2$ (neglecting the $\nu$ mass which is in the sub-eV range).

It is conceivable that the DM particles present in our local 
galactic halo decay inside the cylindrical time projection 
chamber (TPC) of XENON1T detector which has a dimension 
$\sim 1{\rm m}\times 1{\rm m}$~\cite{Aprile:2017aty}.
The photons produced in such decays interact with the electrons 
of the xenon (Xe) atoms via photoelectric effect and 
contribute to the ER events observed in XENON1T. 
If one neglects the $\mathcal{O}$ (eV) ionization energy of Xe, 
the corresponding recoiled electron energy spectrum is a monoenergetic 
peak at the photon energy $E_\gamma$ ($\simeq m_\chi/2$)~\cite{Aprile:2020tmw}
with the event rate ($\mathcal{R}$) given by
%---------------------------------------------------
%\blu{
%\begin{equation}
%\mathcal{R}\, ({\rm{t^{-1} y^{-1}}}) = \frac{1.9 \times 10^{21}}{A}\, \Gamma\, \frac{\rho^{\rm DM}_{\odot}}{m_{\chi}}\, \langle L \rangle\, \sigma_{\rm pe},
%\label{eqn:eventrate}
%\end{equation}
%}
%%\mathcal{R}=\frac{1}{n_T}\, \Gamma\, \frac{\rho^{\rm DM}_{\odot}}{m_\chi}\, \langle L \rangle\, \sigma_{\rm pe}
%% \begin{eqnarray}
%% \mathcal{R}&=&\frac{5.2 \times 10^{15}}{A} \left(\frac{\Gamma}{\rm s^{-1}}\right) \times \\ 
%% && \left(\frac{\rho^{\rm DM}_{\odot}}{\rm GeV/cm^3}\right)
%% \left(\frac{\rm keV}{m_{\chi}}\right) \left(\frac{\langle L \rangle}{\rm m}\right) \left(\frac{\sigma_{\rm pe}}{\rm b}\right)
%% \label{eqn:eventrate}
%%\end{eqnarray}
%where $\rho^{\rm DM}_{\odot} = 0.3$ (in ${\rm GeV}{\rm cm}^{-3}$) is the 
%local energy density of DM particle $\chi$ (assuming $\chi$ accounts for the entire observed DM energy density)~\cite{Bovy:2012tw}, 
%$m_\chi$ is the DM mass (in keV) and $\Gamma$ is the DM 
%decay width (in $\rm s^{-1}$).
%The quantities $A$ ($\approx 131 \rm u$) and $\sigma_{\rm pe}$ (in b) denote the average atomic mass and the photoelectric cross-section 
%of a Xe atom~\cite{osti_4583232}, respectively.
%-------------------------------------------------
\begin{equation}
\mathcal{R}\, = \frac{1}{A}\, \left(\Gamma\, \frac{\rho^{\rm DM}_{\odot}}{m_{\chi}}\, \langle L \rangle\right)\, \sigma_{\rm pe},
\label{eqn:eventrate0}
\end{equation}
where $\rho^{\rm DM}_{\odot} = 0.3$ (in ${\rm GeV}{\rm cm}^{-3}$) 
is the local energy density of DM particle $\chi$ (assuming $\chi$ accounts 
for the entire observed DM energy density)~\cite{Bovy:2012tw}, 
$m_\chi$ is the DM mass (in keV) and $\Gamma$ is the DM 
decay width (in $\rm s^{-1}$).
The quantities $A$ ($\approx 131 \rm u$) and $\sigma_{\rm pe}$ 
(in b) denote the average atomic mass and the photoelectric 
cross-section of a Xe atom~\cite{article,osti_4583232}, 
respectively. The ratio $\rho^{\rm DM}_{\odot}/m_{\chi}$ 
signifies the local number density of DM particles. 
Photons produced from their decays over the average length scale 
$\langle L \rangle$ are capable  of interacting with the Xe 
atoms, where $\langle L \rangle$ (in m) is the mean chord-length 
of the cylindrical TPC of XENON1T. The TPC is considered to be 
a cylinder of length $1{\rm m}$ and diameter $1{\rm m}$~\cite{Aprile:2017aty} 
for which we found $\langle L \rangle \simeq 0.7\,{\rm m}$~\cite{10.2307/3573243}. 
The quantity inside the parenthesis in 
Eq.~\ref{eqn:eventrate0} signifies the flux of the photons 
produced from the DM decays within the XENON1T TPC. In order to 
compare the event rate $\mathcal{R}$  (given in Eq.~\ref{eqn:eventrate0}) 
with that provided by the XENON1T
collaboration~\cite{Aprile:2020tmw,xenon_collaboration_2020_4273099}, 
we have to perform several unit conversions associated with 
$m_\chi$ (keV to GeV), $\Gamma$ (s$^{-1}$ to yr$^{-1}$), 
$\langle L \rangle$ (m to cm) and $\sigma_{\rm pe}$ (b to 
cm$^2$). Additionally, we also need to include the conversion 
factor required to express $\rm{u}^{-1}$ in terms of 
$\rm{t}^{-1}$. One thus ultimately obtains
\begin{equation}
\mathcal{R}\, ({\rm{t^{-1} y^{-1}}}) = \frac{1.9 \times 10^{21}}{A}\, \Gamma\, \frac{\rho^{\rm DM}_{\odot}}{m_{\chi}}\, \langle L \rangle\, \sigma_{\rm pe},
\label{eqn:eventrate}
\end{equation}
where the factor $1.9 \times 10^{21}$ is arising from the 
unit conversions mentioned above.

In principle, decays of DM outside the XENON1T chamber can also 
impart energy to the electrons of the Xe atoms. However, the 
photon in such cases gets mostly shielded. On the other hand, 
neutrinos produced in these decays may enter into the detector 
and lead to electronic recoils via weak interaction. We have 
verified that such effects are down by several orders of 
magnitude, compared to what DM decays within the detector can do 
with the help of photons.

\begin{figure*}[t!]
\centering
\includegraphics[width=8.7cm,height=7.5cm]{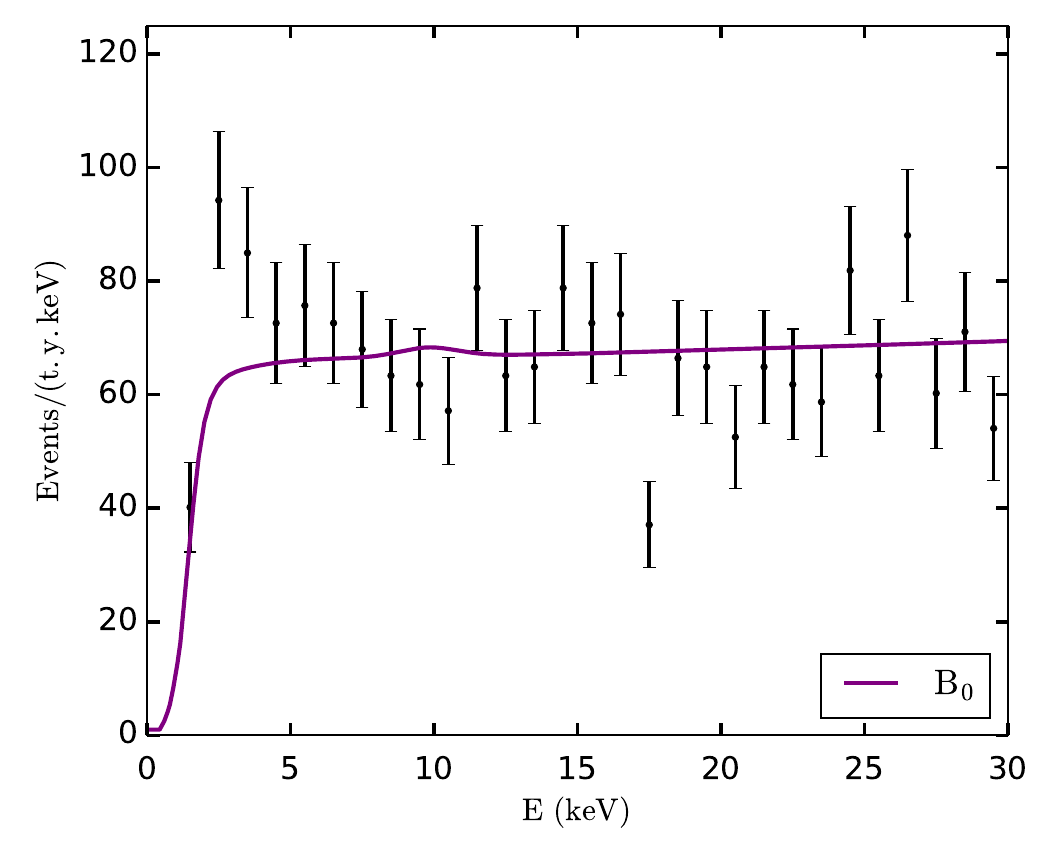}
\includegraphics[width=8.7cm,height=7.5cm]{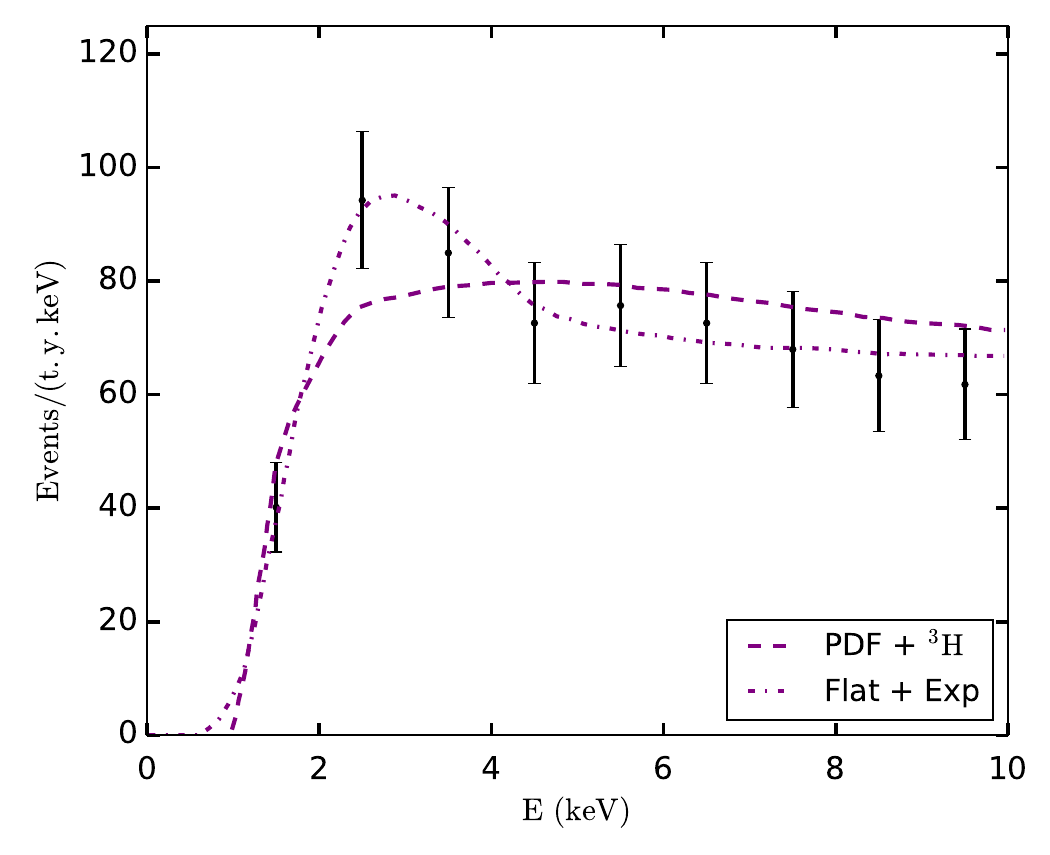}
\caption{
\textit{Left:} The best-fit ER event distribution for the 
standard background $\rm B_0$ (obtained by the XENON1T 
collaboration in
Refs.~\cite{Aprile:2020tmw,xenon_collaboration_2020_4273099})
is shown by the purple solid line. 
\textit{Right:} In case of the two additional backgrounds 
PDF+$\rm ^{3}H$ and Flat+Exp, the corresponding best-fit ER 
event distributions (as available in~\cite{Szydagis:2020isq}) %(taken from~\cite{Szydagis:2020isq}) 
are represented by purple dashed and purple dash-dotted lines, 
respectively. In both of the panels, the XENON1T ER data, along with the 
associated 1$\sigma$ error bars, are shown in
black~\cite{Aprile:2020tmw,xenon_collaboration_2020_4273099}.
%for different background models considered in this work. 
%The XENON1T ER data, along with the 1$\sigma$ error bars, are shown in black~\cite{Aprile:2020tmw,xenon_collaboration_2020_4273099}.
%\red{***  $\rm B_0$ distribution (1-30 keV) in the left panel, PDF$+\rm ^{3}H$ and Flat+Exp (1-10 keV) in the right panel.***}
}
\label{fig:fig0}  
\end{figure*}

In order to obtain the recoiled electron energy spectrum as seen in XENON1T, we convolute the event rate 
given in Eq.~\ref{eqn:eventrate} with a Gaussian distribution with energy resolution
\begin{equation}
\sigma(\rm E)=a\sqrt{\rm E}+b\,\rm E ,
\label{eqn:resol}
\end{equation}
where ${\rm a}=0.31\,\sqrt{\rm keV}$, ${\rm b}=0.0037$~\cite{Aprile:2020tmw} and $\rm E$ (in keV) is the energy of the recoiled electron. 
We have also included the effect of total detector efficiency~\cite{Aprile:2020tmw} in our calculation.  

The primary ER background used in our analysis is the standard 
background $\rm B_0$ considered by the XENON1T 
collaboration~\cite{Aprile:2020tmw,xenon_collaboration_2020_4273099}:
\begin{itemize}
 \item \labelitemi \mbox{\boldmath$\rm B_0$}: 
 $\rm B_0$, consisting of ten distinct background components, 
 represents the most widely used background model employed to fit 
 the ER data recorded by the XENON1T
 experiment~\cite{Aprile:2020tmw, xenon_collaboration_2020_4273099}.
 The $\beta$-decays of $\rm ^{214}Pb$ and $\rm ^{85}Kr$ lead to 
 continuous ER event distributions 
 %, both of which are continuous backgrounds,
 which give most dominant contributions to $\rm B_0$. In 
 addition, monoenergetic peaks coming from the  decays of $\rm 
 ^{131m}Xe$, $\rm ^{125}I$ and $\rm ^{83m}Kr$ are also included 
 along with several other continuous  
 backgrounds~\cite{Aprile:2020tmw}.
\end{itemize}

Additionally, in order to demonstrate how our results vary 
with the choices of the background models, we have taken the 
following backgrounds into account~\footnote{Whether such 
backgrounds indeed contribute to the XENON1T ER data will be 
ultimately decided by future studies.}:
\begin{enumerate}
\item {\bf PDF+\mbox{\boldmath$\rm ^{3}H$}}: 
Presence of tritium ($\rm {^3}H$) within the Xe detector volume  
%can be present within the Xe detector volume either due 
can be attributed to the cosmogenic activation of Xe as well as 
to the atmospheric abundances of tritiated water (HTO) and hydrogen (HT). Though, $\rm {^3}H$ abundances originating from 
cosmogenic activation of Xe and from HTO appear to be negligible, 
the concentration of HT has substantial uncertainty. Furthermore, 
the uncertainties associated with the solubility and diffusion 
properties of $\rm ^{3}H$ within the Xe detector material and 
also the possibilities that tritium may form molecules other than 
HTO and HT make it difficult to rule out $\rm {^3}H$  as a 
possible background component~\cite{Aprile:2020tmw}. In our case,  
$\rm PDF+\rm {^3}H$ refers to the background model where the ER 
event distribution resulting from the $\beta$-decay  
of $\rm ^{3}H$ is added to that of the background mimicking $\rm 
B_0$ (PDF)~\cite{Aprile:2020tmw,xenon_collaboration_2020_4273099, Szydagis:2020isq}. 
The background PDF, considered in our study, is the sum of all the components contributing to $\rm B_0$, 
but with a slightly different energy resolution~\cite{Szydagis:2020isq}, caused by 
the use of a simulation technique different than that used by the XENON1T collaboration~\cite{Aprile:2020tmw}.  
%--------------------------------------------------------
%\blu{
%\item {\bf PDF+\mbox{\boldmath$\rm ^{37}Ar$}}: $\rm ^{37}Ar$ present inside the xenon gas already before filling the detector 
%volume can be an additional source of background~\cite{Aprile:2020tmw}. 
%Decay of $\rm ^{37}Ar$ to the ground state of $\rm ^{37}Cl$ via the electron capturing process produces a 
%2.82 keV monoenergetic photon. $\rm PDF+^{37}Ar$ indicates the background events caused by the 2.82 keV X-ray line of $\rm ^{37}Ar$ added over the background mimicking $\rm B_0$ (PDF)~\cite{Szydagis:2020isq}. 
%}
%--------------------------------------------------------
\item {\bf Flat+Exp}: Flat+Exp depicts a background model 
where a low energy exponential ER distribution is 
added to a flat ER distribution~\cite{Szydagis:2020isq}. 
This background is purely a theoretical one without any 
known physical source but fits the data well 
(see~\cite{Szydagis:2020isq} for details). 
\end{enumerate}

\begin{figure*}[t!]
\centering
\includegraphics[width=12.4cm,height=9.2cm]{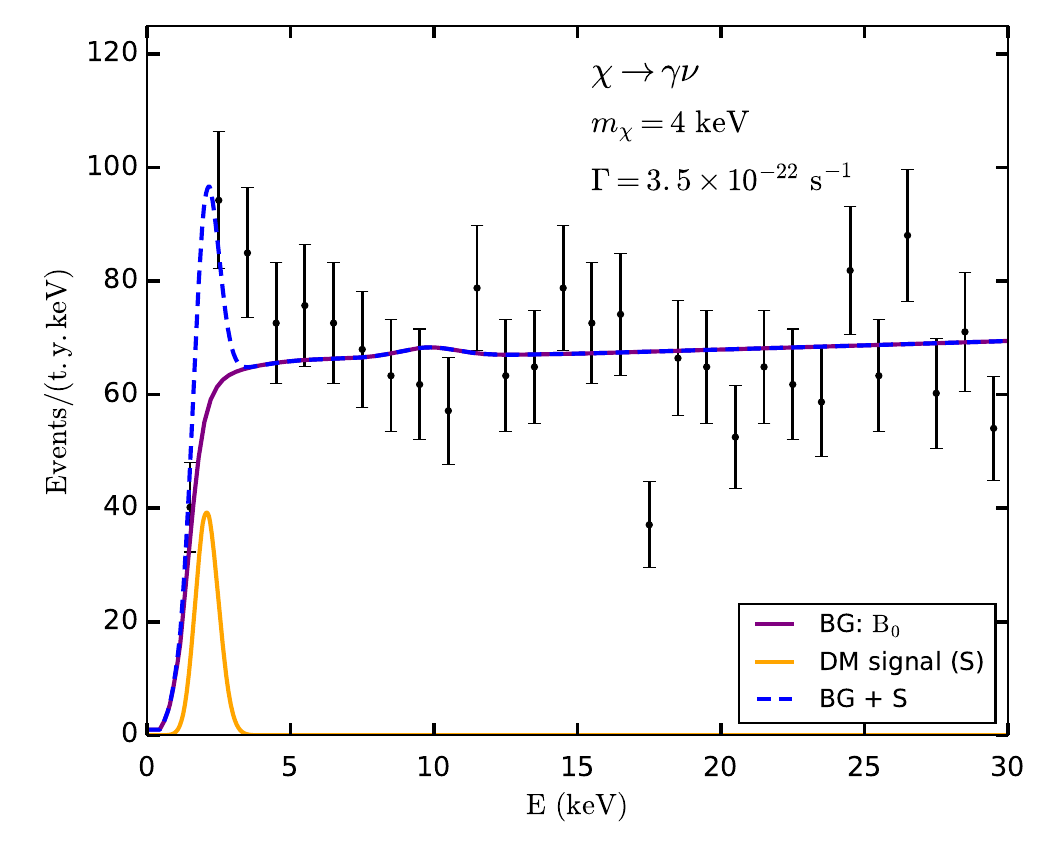}
\caption{
The best-fit event distribution for a DM mass $m_\chi = 4\,{\rm keV}$, considering the standard background $\rm B_0$,  
has been shown as a function of the recoiled electron energy E (blue dashed curve). 
The background (BG) distribution (as given in~\cite{Aprile:2020tmw,xenon_collaboration_2020_4273099}) and the best-fit 
signal rate S (taking the effects of Gaussian smearing and detector efficiency into account) are shown by the 
purple and the orange solid lines, respectively. The XENON1T ER data (along with the error bars) are shown 
in black~\cite{Aprile:2020tmw,xenon_collaboration_2020_4273099}.}
\label{fig:fig1}
\end{figure*}

%In each case, 

%For our primary background $\rm B_0$ (shown in the left panel of 
%Fig.~\ref{fig:fig0}) as well as for the two additional 
%backgrounds (see Fig.~\ref{fig:fig0}; right panel),
%we have considered the best-fit background rates,
%obtained by fitting the corresponding backgrounds against the XENON1T data; 
%see~\cite{Aprile:2020tmw} 
%and \cite{Szydagis:2020isq} for the details of the 
%analyses~\footnote{Only the best-fit rates of each of the 
%backgrounds are available in~\cite{Aprile:2020tmw, xenon_collaboration_2020_4273099} and ~\cite{Szydagis:2020isq}.}.
%In each case, we have considered the best-fit background rates (shown in left and right panel of Fig.~\ref{fig:fig0}) 
%which is 
%obtained by fitting the corresponding background against the XENON1T data; 
%see~\cite{Aprile:2020tmw, Szydagis:2020isq} for details of the analysis~\footnote{Only the best-fit rates 
%of the backgrounds are available in~\cite{Aprile:2020tmw, xenon_collaboration_2020_4273099} and ~\cite{Szydagis:2020isq}.}. 

For our primary background $\rm B_0$ 
as well as for the two additional backgrounds, 
we have considered the best-fit background rates which are shown 
in left and right panels of Fig.~\ref{fig:fig0}, respectively.
These best-fit rates are obtained by fitting the corresponding 
backgrounds against the XENON1T data; 
see~\cite{Aprile:2020tmw} and \cite{Szydagis:2020isq} for the details of the 
analyses~\footnote{Only the best-fit rates of each of the 
backgrounds are available in~\cite{Aprile:2020tmw, xenon_collaboration_2020_4273099} and ~\cite{Szydagis:2020isq}.}.
We then do a binned $\chi^2$ analysis of the observed XENON1T data~\cite{Aprile:2020tmw,xenon_collaboration_2020_4273099} 
with our DM induced signal added to the best-fit rate of each of 
the above-mentioned backgrounds~\cite{Aprile:2020tmw,xenon_collaboration_2020_4273099,Szydagis:2020isq}. 
The $\chi^2$ is defined as follows:
\begin{equation}
\chi^2(\Gamma, m_{\chi})=\underset{i}{\sum}\frac{(n^i_{\rm data}-n^i_{\rm exp})^2}{\sigma^2_i} ,
\label{eqn:chisqdef}
\end{equation} 
where $i$ signifies the energy bin, $n^i_{\rm data}$ is the data
observed in the $i$th bin and $\sigma_i$ is the associated error~\cite{Aprile:2020tmw,xenon_collaboration_2020_4273099}.
$n^i_{\rm exp}$ is the expected number of events in the $i$th bin and is defined as 
$n^i_{\rm exp} = n^i_{\rm background} + n^i_{\rm signal} (\Gamma, m_{\chi})$. 
Here $n^i_{\rm background}$ represents the best-fit background rate for a particular background, as discussed earlier. 
In case of the background $\rm B_0$, the best-fit rate in
~\cite{Aprile:2020tmw,xenon_collaboration_2020_4273099} is 
provided for recoiled electron energy in the range between $1$ keV and $\sim 30$
keV and hence the likelihood analysis is also carried out over 
the same energy range, i.e., first 29 bins of the XENON1T ER 
data~\cite{Aprile:2020tmw,xenon_collaboration_2020_4273099}. 
Due to this reason, the DM mass $m_{\chi}$ (which is twice the 
energy of the produced photon) is varied over the range 
$2 - 60$ keV. 
On the other hand, for the background models PDF+$\rm ^{3}H$ 
and Flat+Exp, the best-fit rates are available up to $\sim 9$ 
keV~\cite{Szydagis:2020isq}, and thus we have performed the 
$\chi^2$ analysis in these cases over the energy range $1$ keV to $\sim 9$ keV (first 9 bins) %of the XENON1T data) 
which well encloses the region of the reported 
excess, i.e., $2 - 3$
keV~\cite{Aprile:2020tmw,xenon_collaboration_2020_4273099}. 
The DM mass $m_\chi$, in these cases, varies
in the range $2 - 18$ keV.
%As a result, for the backgrounds PDF+$\rm ^{3}H$ 
%and Flat+Exp, $m_\chi$ lies in the range $2 - 18$ keV.}   
%------------------------------------------------------------
%For the same reason, \red{in case of the background model $\rm 
%B_0$,} the DM mass $m_{\chi}$ (which is twice the 
%energy of the produced photon) is varied over the range 
%\red{$2 - 60$ keV, while for the backgrounds PDF+$\rm ^{3}H$ 
%and Flat+Exp, $m_\chi$ lies in the range $2 - 18$ keV.}
%------------------------------------------------------------
%\blu{
%Since the best-fit backgrounds in~\cite{Szydagis:2020isq} are available up to $\sim 9$ keV, 
%the $\chi^2$ analysis in our case has been performed over the energy range $1 - \sim 9$ keV (first 9 bins of the XENON1T data)
%which well encloses the region of the reported excess, i.e., $2 - 3$ keV~\cite{Aprile:2020tmw}. 
%For the same reason, the DM mass $m_{\chi}$ (which is twice the energy of the produced photon) is varied over the range $2 - 18$ keV. 
%}
%------------------------------------------------------------
Following~\cite{Bhattacherjee:2020qmv}, for each of the 
three background models, we have done the $\chi^2$ 
minimization over the parameter $\Gamma$ for a DM mass $m_\chi$ 
in the above-mentioned ranges and found the $\chi^2_{\rm min}$ 
which corresponds to the best-fit decay width. The $95\%$ confidence level (C.L.) 
upper and lower limits on the DM decay width are obtained from:
\begin{equation}
\chi^2 = \chi^2_{\rm min} + 2.71 .
\label{eqn:chisq95pdef}
\end{equation}  
In a single parameter fit in terms of $\Gamma$ (for any $m_\chi$), the above $95\%$ C.L. range on both sides of 
$\chi^2_{\rm min}$ should yield the corresponding upper and lower limits on $\Gamma$.

%-------------------------------------------------------------
%\begin{figure*}[ht!]
%\centering
%\includegraphics[width=8.7cm,height=7.5cm]{./Figures/Events_1.pdf}
%\includegraphics[width=8.7cm,height=7.5cm]{./Figures/Events_4_new.pdf}
%\caption{\textit{Left:} The best-fit event distribution for a DM mass $m_\chi = 4\,{\rm keV}$, considering
%the Flat + Exponential background (Flat+Exp), has been shown as a function of the recoiled electron energy E (blue dashed curve). 
%The background (BG) distribution (as given in~\cite{Szydagis:2020isq}) and the best-fit signal rate S 
%(taking the effects of Gaussian smearing and detector efficiency into account) are shown by the 
%purple and orange solid lines, respectively. The XENON1T ER data (along with the error bars) are shown in black~\cite{Aprile:2020tmw}.
%\textit{Right:} Similar set of distributions as shown in the left panel, considering the standard background 
%$\rm B_0$~\cite{Szydagis:2020isq, Aprile:2020tmw}. The BG distribution in this case is shown by the purple dotted line.}
%\label{fig:fig1}
%\end{figure*}
%-------------------------------------------------------------

\section{Astrophysical and Cosmological constraints}
\label{sec:cons}
The DM $\chi$ populates the galaxies, galaxy clusters as well as the extra-galactic continuum. 
Since the DM mass is in the $\mathcal{O}(\rm keV)$ range, the monochromatic X-ray photons produced in the decays 
(i.e., $\chi \rightarrow \gamma \nu$) of such DM populations are possible to be observed in X-ray 
telescopes, e.g., XMM-Newton~\cite{Arviset:2002vq}, Chandra~\cite{Weisskopf:2001uu,Evans_2010}, HEAO-1~\cite{1980ApJ...235....4M,Gruber:1999yr}, 
NuSTAR~\cite{Mori:2015vba, Hong:2016qjq,Lazzarini:2018wlp,Stiele:2018udl}.  
The non-observations of any line like feature in such experiments put constraints in the $\Gamma - m_{\chi}$ plane. 
For any given $m_\chi$, the flux of X-ray photons, originating from the DM decay, is directly proportional to the 
DM decay width $\Gamma$, i.e., larger the decay width is, greater the amount of the expected flux. 
Therefore, by comparing the observed flux with the flux expected from a decaying DM of a particular mass, 
one obtains the upper limit on $\Gamma$ and any higher value of 
$\Gamma$ is thus ruled out for that particular DM mass. 
For example, one can use the XMM observation of X-ray flux from dwarf spheroidal (dSph) galaxies~\cite{Jeltema:2008ax} 
to derive upper limits on $\Gamma$.
We checked that, among all the dSphs observed by XMM, Carina gives the strongest bounds on $\Gamma$.  
The corresponding J-factor (in the field-of-view of XMM) for this dSph has been extracted from~\cite{Geringer-Sameth:2014yza}.
%The J-factor (in the field-of-view of XMM) associated with this dSph has been extracted from~\cite{Geringer-Sameth:2014yza}. 
In parallel, the observations of Andromeda (M31) galaxy by Chandra~\cite{Watson:2011dw} and NuSTAR~\cite{Ng:2019gch}, 
Milky Way (MW) galaxy by Chandra~\cite{Sicilian:2020glg}, Galactic Center (GC) by NuSTAR~\cite{Perez:2016tcq}, 
diffuse cosmic hard X-Ray by HEAO-1~\cite{Essig:2013goa} and 
diffuse X-ray background by XMM and HEAO-1~\cite{Boyarsky:2005us}
%and diffuse X-ray background by XMM~\cite{Boyarsky:2005us} and 
%HEAO-1~\cite{Essig:2013goa} 
also constrain the parameter space of keV DM $\chi$.    
Furthermore, the decay of $\chi$ during the reionization epoch can affect the cosmic microwave background (CMB) power spectrum. 
The observations of CMB temperature and polarization spectra by Planck~\cite{Ade:2015xua} are used to derive upper limits on 
$\Gamma$~\cite{Oldengott:2016yjc}.
It turns out that the observations of Chandra and NuSTAR provide the most stringent upper bounds on $\Gamma$, so far.

\section{Results}
\label{sec:results}

\begin{figure*}[ht!]
\centering
\includegraphics[width=12.4cm,height=9.2cm]{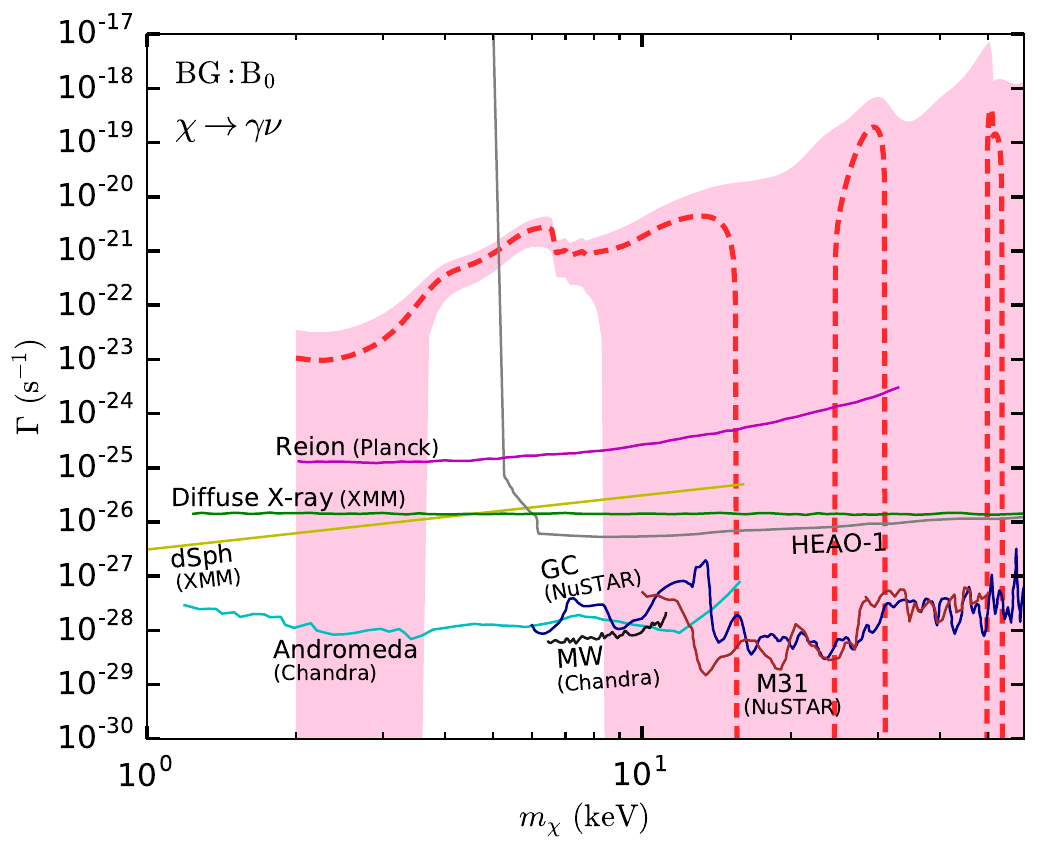}
\caption{
The best-fit DM decay width (red dashed line) and the region 
enclosed by the $95\%$ C.L. upper and lower limits (pink shaded
region), obtained by fitting the XENON1T data~\cite{Aprile:2020tmw,xenon_collaboration_2020_4273099}, 
have been shown as functions of 
the DM mass $m_\chi$, considering the background (BG) model $\rm B_0$ ~\cite{Aprile:2020tmw,xenon_collaboration_2020_4273099}. 
The solid lines represent the upper limits coming from various X-ray observations, e.g., 
XMM observation of dSph Carina (yellow)~\cite{Jeltema:2008ax}, 
%and diffuse X-ray background (green)~\cite{Boyarsky:2005us}, 
HEAO-1 observation of hard diffuse X-ray background (gray)~\cite{Essig:2013goa,Gruber:1999yr}, 
XMM and HEAO-1 observations of diffuse X-ray background (green)~\cite{Boyarsky:2005us},
Chandra observations of Andromeda (cyan)~\cite{Watson:2011dw} and Milky Way(MW) (black)~\cite{Sicilian:2020glg} galaxies, 
NuSTAR observations of Galactic Cente(GC) (blue)~\cite{Perez:2016tcq} and Andromeda(M31) (brown)~\cite{Ng:2019gch} and 
Planck observation of the CMB spectrum (magenta)~\cite{Ade:2015xua,Oldengott:2016yjc} (see the text for details).}
\label{fig:fig2}
\end{figure*}

Fig.~\ref{fig:fig1} shows the best-fit electron recoil energy 
distribution (as a function of the electron energy E) for 
a DM mass $m_\chi = 4\,{\rm keV}$, considering the 
background model $\rm B_0$. In this case, following Eq.~\ref{eqn:chisqdef},
one obtains the best-fit DM decay width $\Gamma = 3.5\times 10^{-22}\,{\rm s}^{-1}$.
The orange solid line, centered around ${\rm E} \simeq m_{\chi}/2 = 2$ keV, 
represents the corresponding signal event distribution 
S (incorporating the effects of Gaussian smearing and detector 
efficiency), while the purple solid line is the event 
distribution for the $\rm B_0$ background (taken 
from~\cite{Aprile:2020tmw,xenon_collaboration_2020_4273099}). 
The blue dashed curve corresponds to the total event distribution 
obtained by adding the signal event distribution (S) to the 
background $\rm B_0$. It can be seen from this figure, 
while $\rm B_0$ fits the data well in the 
higher energy bins, substantial contributions from some new 
physics scenarios (for example, dark matter), are required to 
match the data in the lower energy bins where the actual excess 
was observed. 
 
\begin{figure*}[ht!]
\centering
\includegraphics[width=8.7cm,height=7.5cm]{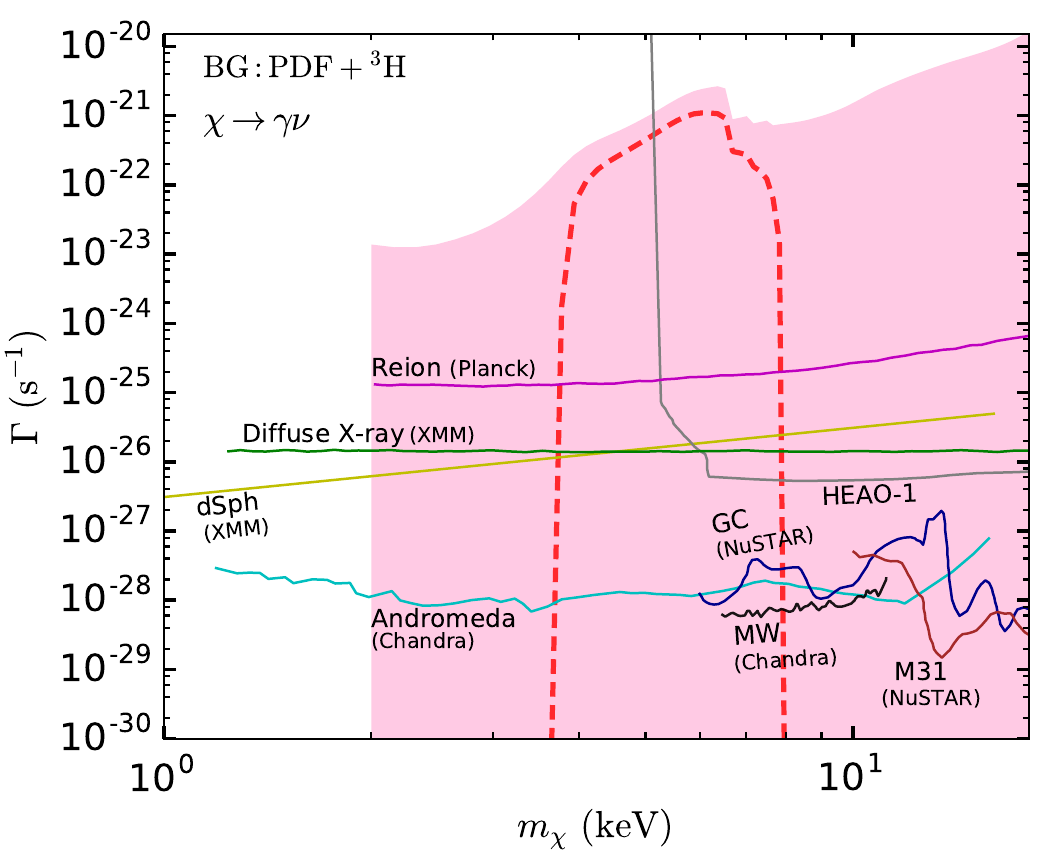}
\includegraphics[width=8.7cm,height=7.5cm]{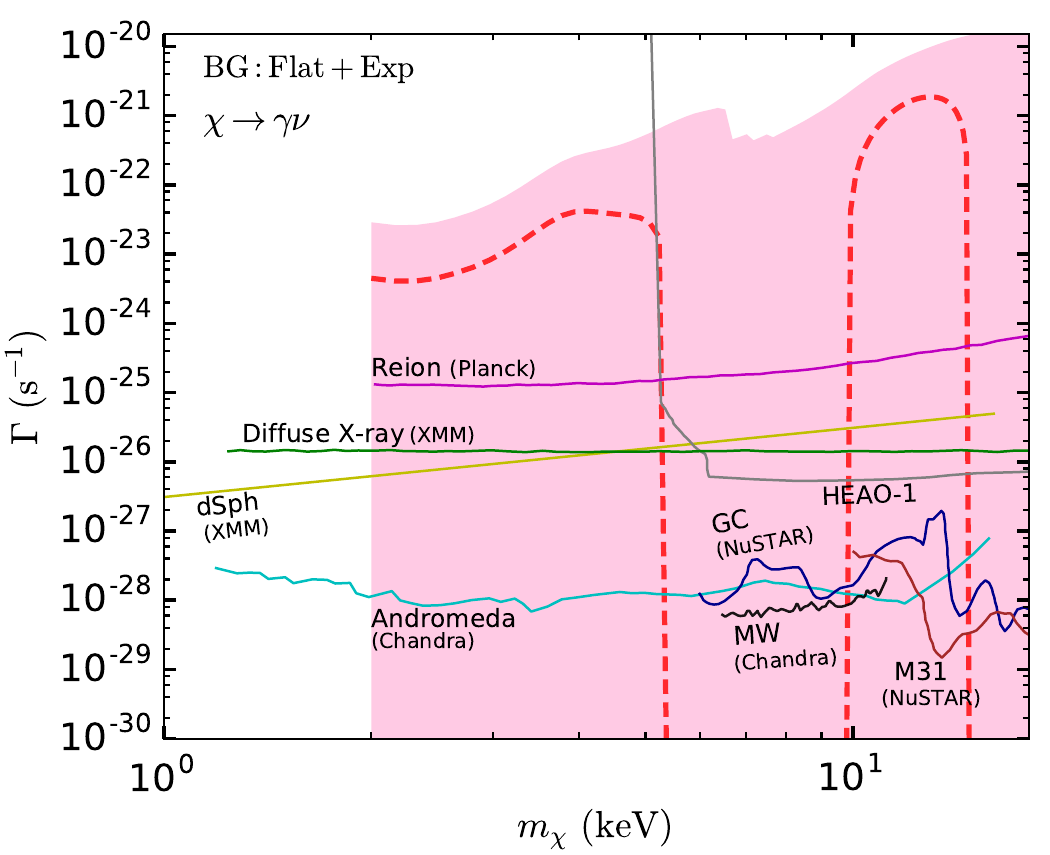}
\caption{
The best-fit $\Gamma$'s (red dashed lines) and the associated 
$95\%$ C.L. bands (pink shaded regions), resulting from the 
analysis of the XENON1T data~\cite{Aprile:2020tmw,xenon_collaboration_2020_4273099}, 
are presented as functions of 
$m_\chi$, taking into account the background (BG) models PDF+$\rm 
^{3}H$ (left panel)~\cite{Szydagis:2020isq} and Flat+Exp (right
panel)~\cite{Szydagis:2020isq}. The astrophysical and cosmological
constraints, shown in each panel, are same as in Fig.~\ref{fig:fig2} (see the text for details).}  
\label{fig:fig3}
\end{figure*}

Fig.~\ref{fig:fig2} represents the constraints on the DM decay width ($\Gamma$), obtained by fitting the XENON1T data, 
for DM mass $m_\chi$ in the range $2 - 60$ keV, considering the background model $\rm B_0$. In addition to the best-fit
$\Gamma$ (dashed line), the area enclosed by the $95\%$ C.L. 
upper and lower limits (shaded region) is also shown.
In the energy range $1.75\,{\rm keV} \lesssim {\rm E} \lesssim 4.25\,{\rm keV}$, the background $\rm B_0$ is much lower than 
the observed excess, even after considering the associated 
2$\sigma$ uncertainty (see left panel of Fig.~\ref{fig:fig0} and 
Fig.~\ref{fig:fig1}) and hence, in order to match the data in 
this energy region, the required DM decay width (i.e., the full 
$95\%$ C.L. band) becomes large for DM masses in the range $3.5 - 8.5$ keV. 
%In the energy range $1.75\,{\rm keV} \lesssim {\rm E} \lesssim 4.25\,{\rm keV}$, the background $\rm B_0$ is far below the 
%$2\sigma$ uncertainty band associated with the observed XENON1T 
%ER data (see left panel of Fig.~\ref{fig:fig0} and 
%Fig.~\ref{fig:fig1}) and hence, in order to match the data in 
%this energy region, both the $95\%$ C.L. lower limit and the 
%best-fit value of $\Gamma$ become large for DM masses in the 
%range $3.5 - 8.5$ keV. 
Outside this energy region, $\rm B_0$ 
always lies within the $2\sigma$ error bars of the data (see  
Fig.~\ref{fig:fig0}; left panel and also Fig.~\ref{fig:fig1}). 
As a result, the $95\%$ C.L. lower limit on 
$\Gamma$ is zero for $2\,{\rm keV} \lesssim m_\chi \lesssim 3.5\,{\rm keV}$ 
and $8.5\,{\rm keV} \lesssim m_\chi \lesssim 60\,{\rm keV}$. The best-fit 
$\Gamma$, on the other hand, is zero only for the DM mass ranges $15\,{\rm keV} \lesssim m_\chi \lesssim 24\,{\rm 
keV}$, $31\,{\rm keV} \lesssim m_\chi \lesssim 50\,{\rm keV}$ and 
$54\,{\rm keV} \lesssim m_\chi \lesssim 60\,{\rm keV}$, since $\rm B_0$ fits 
the data quite well in the corresponding energy domains (see Fig.~\ref{fig:fig0}; left panel and Fig.~\ref{fig:fig1}),
leaving little scope for adding any new signal. 
However, if one thinks in terms of the full 2$\sigma$ uncertainty 
of the XENON1T data, some scope still remains for adding 
new physics signals.
%there always exist some possibilities to add 
%new signals. 
This fact is reflected in the large value of the 
$95\%$ C.L. upper limit on $\Gamma$ throughout the DM mass range 
$2 - 60$ keV.

In Fig.~\ref{fig:fig2}, together 
with the limits obtained from the XENON1T data considering 
$\rm B_0$ as the background model, we have also shown 
the astrophysical and cosmological constraints on $\Gamma$, 
discussed in Sec.~\ref{sec:cons}, by various solid lines. 
The yellow solid line in Fig.~\ref{fig:fig2} represents 
the $95\%$ C.L. upper limit on $\Gamma$, coming from the 
XMM observation of dSph Carina~\cite{Jeltema:2008ax}. 
The upper limits from the observations of diffuse X-ray 
background by XMM and HEAO-1 (at $95\%$ C.L.)~\cite{Boyarsky:2005us}, 
Andromeda (M31) galaxy by Chandra (at $95\%$ C.L.)~\cite{Watson:2011dw} 
and NuSTAR (at $95\%$ C.L.)~\cite{Ng:2019gch}, 
diffuse cosmic hard X-Ray by HEAO-1 (at $95\%$ 
C.L.)~\cite{Essig:2013goa,Gruber:1999yr}, 
Milky Way (MW) galaxy by Chandra (at $99\%$ 
C.L.)~\cite{Sicilian:2020glg}, 
Galactic Center (GC) by NuSTAR (at $95\%$ 
C.L.)~\cite{Perez:2016tcq} are shown by 
green, cyan, brown, gray, black and blue solid lines, respectively.
The Planck CMB constraint on $\Gamma$ (at $95\%$ C.L.) 
has been shown in Fig.~\ref{fig:fig2} by magenta solid line~\cite{Oldengott:2016yjc}.
As mentioned earlier, the Chandra observations of Andromeda and Milky Way (MW) galaxies 
as well as the NuSTAR observations of Andromeda (M31) 
and Galactic Center (GC) put the tightest constraints in the $\Gamma - m_{\chi}$ plane.

It can be seen from Fig.~\ref{fig:fig2} that for the 
background model $\rm B_0$, the $95\%$ C.L. upper limit on 
$\Gamma$, obtained by analyzing the XENON1T data, is ruled out by 
X-ray observations as well as by Planck CMB observation over the mass range $2 - 60$ keV. 
However, if one considers the full $95\%$ C.L. band 
associated with the best-fit $\Gamma$, %including the best-fit, 
there still exist some scopes 
to allow the DM induced signals. For example, 
since the $95\%$ C.L. lower limit on $\Gamma$ is zero for 
$2\,{\rm keV} \lesssim m_\chi \lesssim 3.5\,{\rm keV}$ and 
$8.5\,{\rm keV} \lesssim m_\chi \lesssim 60\,{\rm keV}$, 
any finite $\Gamma$ in these mass ranges which is lower than 
all the astrophysical constraints is allowed at $95\%$ C.L. 
On the other hand, for DM masses in the range 
$3.5\,{\rm keV} \lesssim m_\chi \lesssim 8.5\,{\rm keV}$, 
the full $95\%$ C.L. band is disfavoured.

The best-fit $\Gamma$'s and the associated $95\%$ C.L. bands, 
resulting from the analysis of the XENON1T data using 
background models PDF+$\rm ^{3}H$ and Flat+Exp are shown in 
left and right panels of Fig.~\ref{fig:fig3}, respectively.
Both of these backgrounds are well inside the 2$\sigma$ error bars 
of the XENON1T data in all energy bins 
(see Fig.~\ref{fig:fig0}; right panel) and thus, in each case, the
$95\%$ C.L. lower limit on $\Gamma$ is zero over the entire DM 
mass range considered, i.e., $2 - 18$ keV. Although the $95\%$ 
C.L. lower limits on $\Gamma$, for these two backgrounds, show 
similar behaviour throughout the mass domain $2 - 18$ keV, the 
best-fit $\Gamma$'s are significantly different. 
For the PDF+$\rm ^{3}H$ background, the best-fit $\Gamma = 0$ for 
$2\,{\rm keV} \lesssim m_\chi \lesssim 3.5\,{\rm keV}$ and 
$ 8\,{\rm keV} \lesssim m_\chi \lesssim 18\,{\rm 
keV}$, while it is relatively larger for other DM masses because 
the background falls below 
%falls below the 1$\sigma$ error bars of 
the data in the corresponding energy regions (see right panel of 
Fig.~\ref{fig:fig0}). In case of Flat+Exp background, the 
best-fit $\Gamma$ is zero for $5\,{\rm keV} \lesssim m_\chi 
\lesssim 10\,{\rm keV}$ and $15\,{\rm keV} \lesssim m_\chi 
\lesssim 18\,{\rm keV}$, since this background fits the data well 
in the energy ranges $2.5\,{\rm keV} \lesssim {\rm E} \lesssim 
5\,{\rm keV}$ and $7.5\,{\rm keV} \lesssim {\rm E} \lesssim 
9\,{\rm keV}$ (see Fig.~\ref{fig:fig0}; right panel).
Like in the case of $\rm B_0$, for PDF+$\rm ^{3}H$ 
and Flat+Exp backgrounds, too, the $95\%$ C.L. upper limits on 
$\Gamma$ are large for all %the 
DM masses in the range $2 - 18$ keV. Along with the XENON1T 
results, similar astrophysical and cosmological constraints, 
shown in Fig.~\ref{fig:fig2}, are also presented in each panel of 
Fig.~\ref{fig:fig3}. By comparing these constraints with those 
obtained from the analysis of the XENON1T data, one finds that, 
though the $95\%$ C.L. upper limits on $\Gamma$ are always ruled 
out for both of the background models, in each case, a 
significant portion of the DM parameter space in the $95\%$ 
C.L. region, which is consistent with all the astrophysical 
observations, exists for all DM masses in the range $2 - 18$ keV.
This observation is in contrast to what has been seen in the
case of $\rm B_0$.

The local DM density $\rho^{\rm DM}_{\odot}$ is an important 
quantity in determining the ER event rate in the XENON1T 
experiment (see Eq.~\ref{eqn:eventrate0} and 
Eq.~\ref{eqn:eventrate}). In this study we have considered the 
local DM density $\rho^{\rm DM}_{\odot}$ 
to be fixed at its central value $0.3\,{\rm GeV}{\rm cm}^{-3}$~\cite{Bovy:2012tw}. If one considers the full $1\sigma$ range of 
$\rho^{\rm DM}_{\odot} = 0.2-0.4 \,{\rm GeV}{\rm cm}^{-3}$~\cite{Bovy:2012tw}, all finite constraints on $\Gamma$ 
(both the best-fit and the $95\%$ C.L.) will vary by a factor of $0.7-1.3$, i.e., within $30\%$ of the values presented here.

\section{Summary and conclusions}
\label{sec:conclusion}

We have studied, in the context of the recent XENON1T 
observation, a fermionic keV DM scenario where the DM 
particle $\chi$ decays into a photon and a SM neutrino. 
Photons produced in the decay of local population of these DM 
particles inside the XENON1T chamber are absorbed by the 
electrons of the Xe atoms and thereby contribute to the 
recoiled electron events recorded by XENON1T. 
We have primarily considered the standard background 
$\rm B_0$~\cite{Aprile:2020tmw,xenon_collaboration_2020_4273099} 
for our study. %in our analysis.}
%We have considered the standard background $\rm B_0$~\cite{Aprile:2020tmw,xenon_collaboration_2020_4273099}
%as the primary background for our study. %analysis. 
By adding $\rm B_0$ to the DM induced signal and performing an 
one parameter $\chi^2$ analysis against the XENON1T data, 
we have obtained the best-fit values of the 
DM decay width $\Gamma$ and the associated $95\%$ C.L. band as 
%a function 
functions of the DM mass $m_\chi$ in the range $2 - 60$ 
keV. In addition, two other background models, namely, 
PDF+$^{3} \rm H$~\cite{Szydagis:2020isq} and
Flat+Exp~\cite{Szydagis:2020isq} are also taken into account, 
for the sake of illustration. 
Although PDF+$^{3} \rm H$ has substantial uncertainty 
associated with it and Flat+Exp is purely speculative,
they provide comparatively better fit to the observed XENON1T 
excess. For both of these backgrounds, the corresponding  
best-fit $\Gamma$'s and the $95\%$ C.L. bands are 
estimated for the DM mass range $2 - 18$ keV, by carrying out an 
analysis similar to the case of $\rm B_0$.

On the other hand, photons produced in the decays of keV DM particles, 
occurring in the extra-galactic continuum and various 
astrophysical structures, are expected to be observed in X-ray observations. 
These keV DM particles can also decay during the reionization epoch and 
leave footprints in the CMB spectrum observed by Planck. 
Constraints coming from these observations in the $\Gamma-m_\chi$ plane 
have been taken into account and juxtaposed 
with the results obtained from the XENON1T data. 
We find that, the DM mass range $3.5\,{\rm keV}\lesssim m_\chi 
\lesssim 8.5\,{\rm keV}$ is strongly disfavoured at 95$\%$ C.L. 
by all astrophysical observations so long as the most commonly 
used background $\rm B_0$ is considered to be the only reliable 
source of background. However, for DM masses outside the 
above-mentioned window  (i.e., $2\,{\rm keV}\lesssim m_\chi 
\lesssim 3.5\,{\rm keV}$ and $8.5\,{\rm keV}\lesssim m_\chi 
\lesssim 60\,{\rm keV}$), substantial parts of the DM parameter 
space are still allowed at 95$\%$ C.L. Such conclusions regarding 
the allowed regions in the DM parameter space depend on the 
energy distribution of the background model under consideration, 
a fact which is demonstrated by choosing the backgrounds 
PDF+$^{3} \rm H$ and Flat+Exp. In both of these cases, reasonable 
portions of the DM parameter space that lie within the $95\%$ 
C.L. bands throughout the DM mass range $2 - 18$ keV are found to 
be consistent with all astrophysical data.

%In case of the backgrounds PDF+$^{3} \rm H$ 
%and Flat+Exp, reasonable portions of the DM parameter space that 
%lie within the $95\%$ C.L. bands over the DM mass range 
%$2 - 18$ keV, are found to be consistent with all astrophysical data.
%----------------------------------------------------------
%\blu{
%We find that, for each of the three background models Flat+Exp, PDF+$^{37} \rm Ar$ and 
%PDF+$^{3} \rm H$, a reasonable portion of the DM parameter space that lies within the $95\%$ C.L. band over the entire mass range considered
%is allowed by all astrophysical observations.
%Considering the background $\rm B_0$, the allowed parameter space in the $95\%$ C.L. band spans most of the mass range, 
%except for a region $3 - 8$ keV, where the full $95\%$ C.L. band is disallowed by astrophysical data.
%}
%----------------------------------------------------------

The results presented in Figs.~\ref{fig:fig1}, ~\ref{fig:fig2} 
and ~\ref{fig:fig3} can be appropriately translated in terms of 
the parameters of specific theoretical models, including 
those mentioned in the introduction. The R-parity violating 
gravitino DM scenario is in general quite constrained in this 
respect, since the `mixing parameter' driving the DM decay is 
subjected to tight limits from neutrino masses on the one hand, 
and restrictions on the inter-connected particle spectrum on the 
other. The axino DM scenario is less restricted, being possessed 
with relatively larger number of phenomenological parameters.

As mentioned in the introduction, the constraints obtained from 
the XENON1T data as well as the X-ray observations etc., assume 
that the DM density of our universe is saturated by warm DM. 
For the $m_\chi$-range of our interest, such saturation can be 
questionable if the constraints from the Lyman-$\alpha$ 
forest are imposed, as has been claimed, for example, 
in~\cite{Irsic:2017ixq}. Higher values of $\Gamma$ will be 
required in such cases, leading to some quantitative changes in 
all the constraints. For example, if a maximum 
fraction $f$ of the relic density can be attributed to warm DM, 
then the limits on $\Gamma$ from every considerations in 
Fig.~\ref{fig:fig2} and in each panel of Fig.~\ref{fig:fig3} get 
scaled up to $\Gamma/f$. 

To conclude, the recent XENON1T data may be interpreted as the 
source of additional constraints on dark matter scenarios, 
the severity of the constraints being dependent on detailed 
understanding of the backgrounds. This become evident from the 
above study on otherwise allowed decaying warm dark matter.  

\begin{acknowledgments}
The authors thank Pijushpani Bhattacharjee and 
Subinoy Das for useful discussions.
The work of KD was partially  supported by the grant MTR/2019/000395, funded by SERB, DST, Government of India. 
The work of AG and AK was partially supported by funding available from the Department of Atomic Energy,
Government of India, for the Regional Centre for Accelerator-based Particle Physics (RECAPP),
Harish-Chandra Research Institute, Allahabad. 
\end{acknowledgments}

%************************************************************
\bibliographystyle{apsrev4-1}
\bibliography{refs}

\end{document}